\documentclass[aps,prl,reprint,superscriptaddress,twocolumn,showpacs]{revtex4-1}
\usepackage[dvipdfmx]{graphicx}
\usepackage[version=3]{mhchem}
\usepackage{braket}
\usepackage{natbib}
\usepackage{here}
\usepackage{dcolumn}
\usepackage{color}
\usepackage{bm}
\newcommand{\AffISSP}{\affiliation{Institute for Solid State Physics, University of Tokyo, Kashiwanoha, Chiba 277-8581, Japan}}
\newcommand{\AffUTPhys}{\affiliation{Department of Physics, University of Tokyo, Hongo, Tokyo 113-0033, Japan}}
\newcommand{\AffJASRI}{\affiliation{Japan Synchrotron Radiation Research Institute, 1-1-1 Kouto, Sayo, Hyogo 679-5198, Japan}}
\newcommand{\AffIMR}{\affiliation{Institute for Materials Research, Tohoku University, Sendai, Miyagi 980-8577, Japan}}
\newcommand{\AffUH}{\affiliation{Graduate School of Material Science, University of Hyogo, Kamigori, Hyogo 678-1297, Japan}}
\newcommand{\AffRIKEN}{\affiliation{RIKEN SPring-8 Center, 1-1-1 Kouto, Sayo, Hyogo 679-5148, Japan}}
\newcommand{\AffIMS}{\affiliation{Institute for Molecular Science, Okazaki, Aichi 444-8585, Japan}}
\newcommand{\AffIMSS}{\affiliation{Institute of Materials Structure Science, High Energy Accelerator Research Organization, Tsukuba, Ibaraki 305-0801, Japan}}
\newcommand{\AffCSRN}{\affiliation{Center for Spintronics Research Network, Tohoku University, Sendai, Miyagi 980-8577, Japan}}
\usepackage{amsmath}	
\begin{document}
\title{Ultrafast demagnetization of Pt magnetic moment in L1${}_0$-FePt probed by hard x-ray free electron laser}
\author{Kohei~Yamamoto}
\homepage{http://sites.google.com/site/yamakolux/}
\email{yamako@issp.u-tokyo.ac.jp}\AffISSP\AffUTPhys
\author{Yuya~Kubota}\AffJASRI
\author{Motohiro~Suzuki}\AffJASRI
\author{Yasuyuki~Hirata}\AffISSP\AffUTPhys
\author{Kou~Takubo}\AffISSP
\author{Yohei~Uemura}\AffIMS
\author{Ryo~Fukaya}\AffIMSS
\author{Kenta~Tanaka}\AffUH
\author{Wataru~Nishimura}\AffUH
\author{Takuo~Ohkochi}\AffJASRI
\author{Tetsuo~Katayama}\AffJASRI
\author{Tadashi~Togashi}\AffJASRI
\author{Kenji~Tamasaku}\AffRIKEN
\author{Makina~Yabashi}\AffJASRI\AffRIKEN
\author{Yoshihito~Tanaka}\AffUH
\author{Takeshi~Seki}\AffIMR\AffCSRN
\author{Koki~Takanashi}\AffIMR\AffCSRN
\author{Hiroki~Wadati}\AffISSP\AffUTPhys
\date{\today}
\begin{abstract}
We demonstrate ultrafast magnetization dynamics in a 5d transition metal using circularly-polarized x-ray free electron laser in the hard x-ray region. 
A decay time of light-induced demagnetization of L1${}_0$-FePt was determined to be $\tau_\textrm{Pt} = 0.6\ \textrm{ps}$ using time-resolved x-ray magnetic circular dichroism at the Pt L${}_3$ edge, whereas magneto-optical Kerr measurements indicated the decay time for total magnetization as $\tau_\textrm{total} = 0.1\ \textrm{ps}$. 
A transient magnetic state with the photo-modulated magnetic coupling between the 3d and 5d elements is firstly demonstrated.
\end{abstract}
\maketitle


\begin{figure*}
\begin{center}
  \includegraphics[clip,width=18cm]{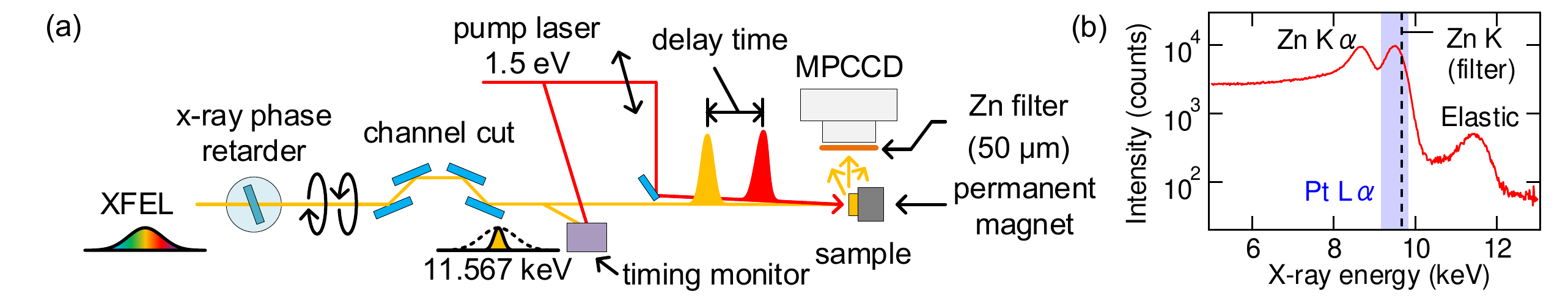}
  \caption{(a) Experimental setup for trXMCD constructed at SACLA BL3.
(b) Fluorescence spectrum.
The shaded area indicates the energy window of the Pt L$\alpha$ emission line.}
  \label{fig:setup}
  \end{center}
\end{figure*}

In recent times, photo-induced magnetism has attracted significant attention from researchers because of its non-trivial mechanism and avenues for practical application~\cite{Kirilyuk2010}. 
An early study on photo-induced magnetism involved ultrafast demagnetization of ferromagnetic Ni foil within 1 ps studied by the magneto-optic Kerr effect (MOKE)~\cite{Beaurepaire1996}.
Since then, extensive investigations have been conducted to find more novel photo-induced magnetic phenomena and explain their mechanisms~\cite{Stanciu2007,Mangin2014,Kimel2014,Lambert2014};
among these, the phenomenon of all-optical helicity-dependent switching (AOHDS) has been a hot topic for research.
Ferrimagnetic GdFeCo was reported to show magnetization switching when it was exposed to a circular polarized optical laser~\cite{Stanciu2007};
subsequent studies have revealed that a wide range of ferrimagnetic materials, including TbCo, DyCo, and HoFeCo, show similar AOHDS phenomena~\cite{Mangin2014}.
Till recently, AOHDS was considered to be possible only in ferrimagnetic thin films that has a compensation temperature near the room temperature with perpendicular magnetic anisotropy (PMA)~\cite{Kimel2014};
however, it was observed that ferromagnetic Co/Pt superlattice thin films and granular L1${}_0$-FePt also exhibited AOHDS~\cite{Lambert2014}. Hence, thus far, the necessary conditions required for observing AOHDS have not been established.

Considering this, to understand the mechanism of photo-induced magnetization dynamics, element-specific measurements are necessary because magnetic materials that show remarkable photo-induced behaviors contain more than one magnetic element~\cite{Kimel2014}.
X-ray magnetic circular dichroism (XMCD) is an element-specific magnetism measurement technique~\cite{Stohr2006,Chen1995} and time-resolved x-ray magnetic circular dichroism (trXMCD) can provide information on element-specific dynamics of magnetic materials.
In a previous soft x-ray trXMCD study at Fe~L and Gd~M edges, it was shown that magnetization dynamics show a transient ($\sim 1~\textrm{ps}$) ferromagnetic state in ferrimagnetic GdFeCo by photo-excitation; in particular, it was considered that this ferromagnetic state drives magnetization reversal via magnetization coupling modulated by laser illumination~\cite{Radu2011}.

It is also important to directly observe the dynamics of Pt magnetic moment using an element-specific experimental probe such as trXMCD because it can be used to analyze hard magnetic materials with PMA exhbiting AOHDS containing Pt.
For trXMCD, soft x-rays have been primarily used on the L (2p$\rightarrow$3d) and M (3d$\rightarrow$4f) edges of 3d and 4f elements, respectively.
Nevertheless, to the best of our knowledge, the L (2p$\rightarrow$5d) edge of 5d transition metals, which exists in the hard x-ray region, has not been used for trXMCD; thus, element-specific spin dynamics of ferromagnetic materials with Pt have not been clarified.
In another study, ultraviolet light from a high harmonic generation laser was applied on the O edge (5p$\rightarrow$5d) of Pt and M edge of Co to observe trXMCD~\cite{Willems2015}; however, the XMCD of the O edge of Pt was not sufficiently large to determine the precise time scale for Pt demagnetization.
Owing to their large spin-orbit interaction, 5d elements play an important role in magnetism; furthermore, trXMCD at L edge in the hard x-ray region is necessary to entirely understand the ultrafast demagnetization process. 
X-ray free electron laser (XFEL) enabled time-resolved measurements of ultrafast phenomena with its ultrashort x-ray pulses with a duration of $< 1\ \textrm{ps}$~\cite{Emma2010,Ishikawa2012} and it has been applied to a lot of research fields including magnetization dynamics~\cite{Malvestuto2018}.
In this study, we report the first trXMCD measurements in the hard x-ray region using the XFEL, for which an x-ray phase retarder was used to produce circular polarized x-rays~\cite{Suzuki2014}.
We successfully determined the photo-induced ultrafast spin dynamics of Pt magnetic moments in L1${}_0$-FePt thin films with PMA.

We used single crystal thin films of L1${}_0$-FePt alloy with a thickness of 20 nm as samples for our experiment.
These samples were fabricated on MgO(100) substrates epitaxially using the sputter method at a substrate temperature of 500${}^\circ$C. 

Figure~\ref{fig:setup}(a) shows the experimental setup for the pump-probe trXMCD based on the fluorescence yield method in SACLA BL3~\cite{Ishikawa2012}.
In this beamline~\cite{Tono2013}, time-resolved hard x-ray absorption spectroscopy~\cite{Katayama2013,Uemura2016} and diffraction~\cite{Hartley2017,TANAKA2013,Newton2014} have been reported; however, thus far, time-resolved x-ray absorption measurements with circular polarized XFEL pulses have not been performed.
To combine the diamond x-ray phase retarder~\cite{Suzuki2014} and the timing-monitor system~\cite{Beye2012,Sato2015,Katayama2016}, we adopted the optics setup of somewhat acrobatic; a pink (quasi-monochromatic)  self-amplified spontaneous emission radiation with the bandwidth of $\sim 40$ eV illuminated the diamond crystal tuned to generate circular polarization at a target x-ray energy, and the following four-bounce channel-cut monochromator selected the monochromatic beam with circular polarization.
We confirmed that the setup with a diamond crystal with the thickness of 1.5 mm and Si 111 channel-cut crystals worked well to produce circular polarization with a high degree of circular polarization of $P_\textrm{c} > 0.9$ at 11.567 keV~\footnote{Y.~Kubota \textit{et al.}, in preparation}.
The timing monitor system~\cite{Beye2012,Sato2015,Katayama2016} effectively compensated the jitter of XFEL pulses and the resulting temporal resolution blow 50 fs was achieved.

X-ray fluorescence detection method was used to acquire x-ray absorption and XMCD signals using a multi-port charge-coupled device (MPCCD) detector~\cite{Kameshima2014}.
The circularly polarized x-ray beam at the Pt L${}_3$ edge ($\sim$~11.6 keV) was incident on the sample in the direction normal to the film plane at a repetition rate of 30 Hz.
X-ray fluorescence emitted from the sample was photon-counted and energy-analyzed by the MPCCD placed at the side with a distance of 150 mm.
A Zn filter with a thickness of 50 $\mu\textrm{m}$ was used to eliminate the strong elastic and Compton scattering from the sample substrate.
Figure~\ref{fig:setup}(b) shows a typical x-ray fluorescence spectrum, in which the Pt L$\alpha$ fluorescence was observed around 9.4~keV.
X-ray absorption signals of Pt were measured as the photon counts integrated over an energy window of 9.15-9.83 keV.
The magnetization of the sample was saturated with a perpendicular magnetic field of 0.6 T applied using a permanent magnet.
A Ti:Sapphire laser with a photon energy of 1.5~eV with the pulse duration of approximately 30 fs was used to pump.

\begin{figure}
\begin{center}
  \includegraphics[clip,width=8cm]{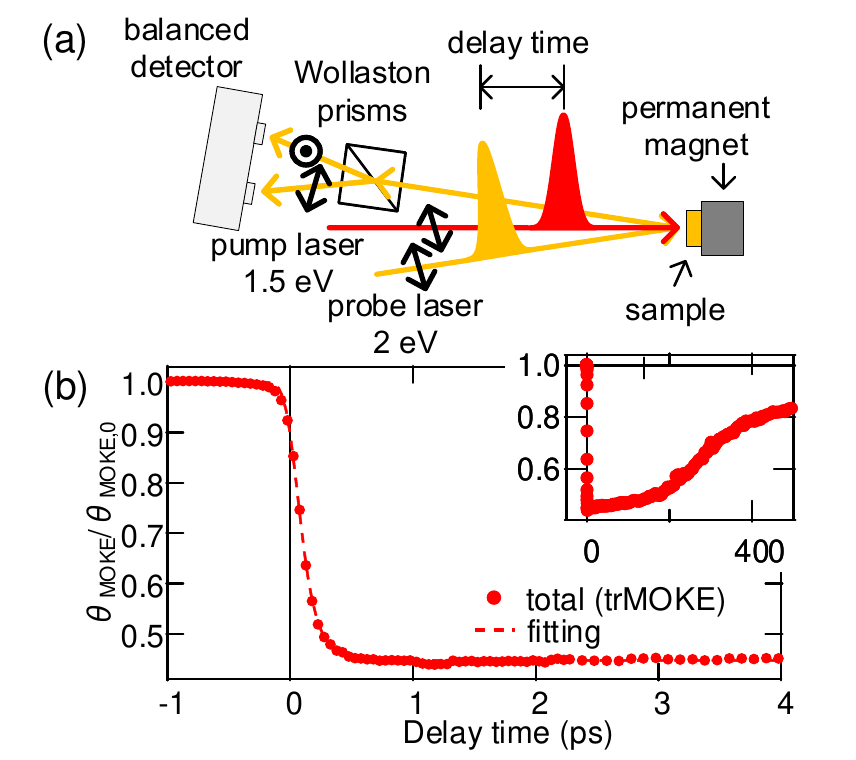}
  \caption{TrMOKE of L1${}_0$-FePt thin films.
(a) Experimental setup used for trMOKE measurements. (b) Experimental results with a fluence of $32\ \textrm{mJ/cm}^2$ as well as exponential fitting curve .
}
\label{fig:trKerr}
  \end{center}
\end{figure}

To characterize the dynamics of total magnetization of the sample, time-resolved MOKE (trMOKE) measurements were performed using visible laser as pump and probe light with the setup shown in Fig.~\ref{fig:trKerr}(a).
A Ti:Sapphire regenerative amplifier with a photon energy of 1.5 eV, a pulse width of 100 fs, and a repetition rate of 1~kHz was used as the light source. 
The pump beam was a fundamental light, and its repetition rate was decreased to the half of the fundamental frequency (500~Hz) with an optical chopper.
The probe beam with a fundamental frequency of 1~kHz had a photon energy of 2 eV which converted by a optical parametric amplifier, and was incident nearly normally on the film surface, after which the polarization of the reflected light was analyzed.

Figure~\ref{fig:trKerr}(b) shows trMOKE results of L1${}_0$-FePt with the pump laser fluence of $32~\textrm{mJ}/\textrm{cm}^2$.
The Kerr rotation angle $\theta_{\textrm{MOKE}}$ normalized by an angle before excitation $\theta_{\textrm{MOKE},0}$, reduces to about 40 \% of that in the unexcited state.
The charactoristic demagnetization time $\tau_\textrm{total}$ was calculated via fitting using a single exponential function:$f(t)=0\ (t \le 0)$ and $a\left[1-\exp(-t/\tau_\textrm{total})\right]\ (t>0)$, convoluted by gaussian function.
We determined that the time scale is $\tau_\textrm{total}=0.1~\textrm{ps}$ when the same pump laser fluence as trXMCD.
The time constant $\tau_\textrm{total}$ is comparable to that in the previous trMOKE study, i.e., 0.15-0.38 ps~\cite{Mendil2014}.

\begin{figure}
\begin{center}
  \includegraphics[clip,width=8cm]{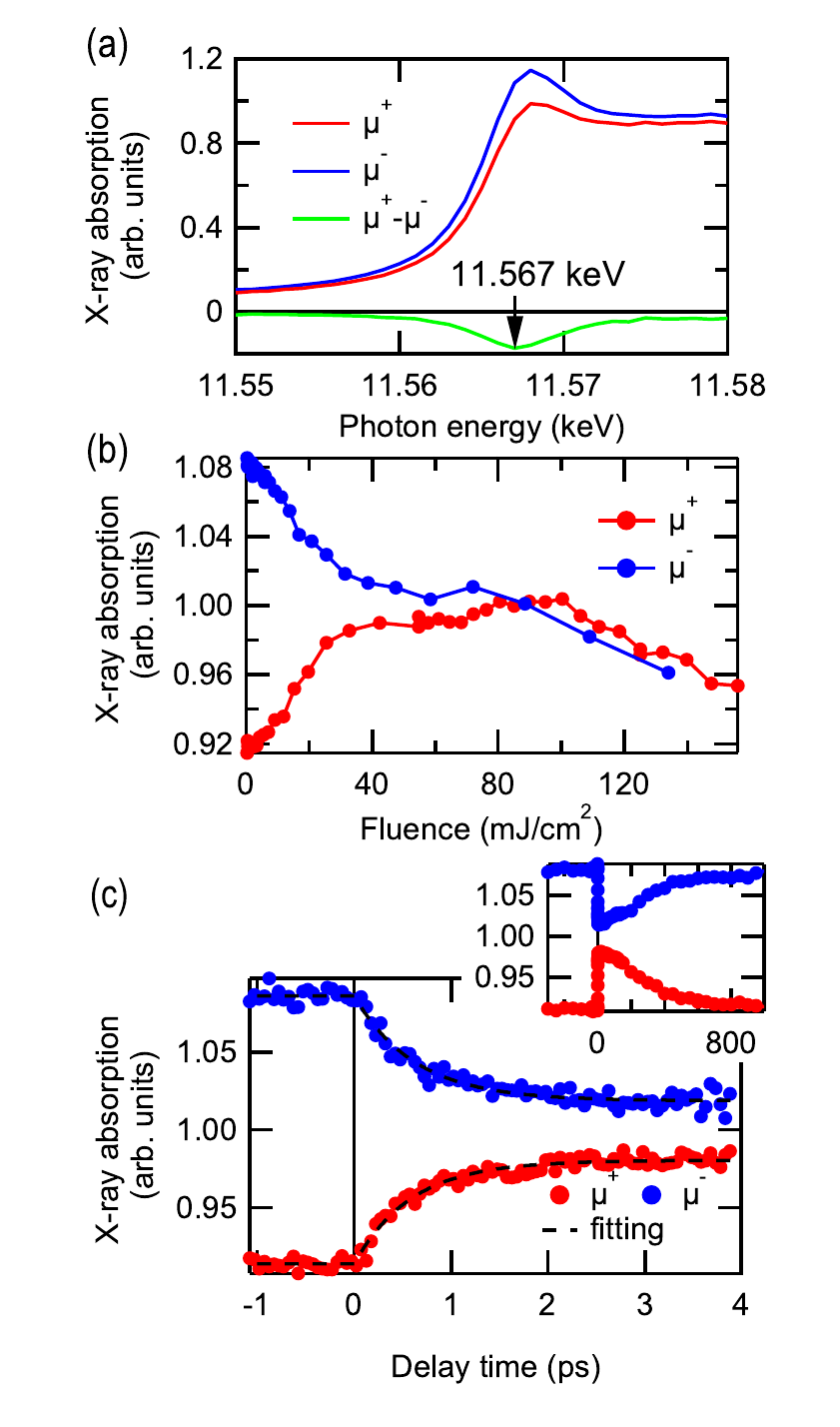}
  \caption{TrXMCD of L1${}_0$-FePt thin films. (a) Static x-ray absorption spectra at the Pt L${}_3$ edge. The photon energy was determined to be 11.567 keV as indicated by the arrow.
  (b) Transient XMCD intensity plot with a delay of 100 ps as a function of laser fluence. 
(c) Delay scan graphs of the x-ray absorption intensity. The inset shows the result over a wide time range.
}
  \label{fig:fluspec}
  \end{center}
\end{figure}

First we show the results of x-ray absorption and the XMCD spectrum in L1${}_0$-FePt using our setup in SACLA.
Figure~\ref{fig:fluspec}(a) shows the x-ray absorption at the Pt L${}_3$ edge, measured for positive ($\mu^+$) and negative ($\mu^-$) photon helicities, respectively.
$\mu^+$ and $\mu^-$ are normalized to unity such that the polarization-averaged XAS is unity, i.e., $(\mu^++\mu^-)/2=1$, at 11.567~keV before excitation.
XMCD is defined as the difference of x-ray absorption of different helicities, $\Delta\mu=\mu^+-\mu^-$.
The clear XMCD spectra with the magnitude of $\sim 15\%$ with respect to the edge jump was observed and are in good agreement with the results using a synchrotron beamline~\cite{Ikeda2017}.

TrXMCD measurements were performed at an x-ray energy of 11.567 keV, at which the Pt XMCD spectrum takes the maximum.
Figure~\ref{fig:fluspec}(b) shows the dependence of the trXMCD amplitude on the fluence of the pump laser at a fixed delay time of 100 ps.
With the fluence of 40-80 $\rm{mJ/cm}^2$, the XMCD intensity is almost constant.
Above 80 $\rm{mJ/cm}^2$, both $\mu^+$ and $\mu^-$ decreased simultaneously, which was a signature of a destructive damage of the sample due to the strong irradiation of the optical laser.
Therefore, we set the laser fluence to 32 $\rm{mJ/cm}^2$ to avoid damaging the sample and this condition was therefore used to obtain the trXMCD data presented in the following.
The results of pump-probe delay scan are shown in Fig.~\ref{fig:fluspec}(c).
X-ray absorption signals of $\mu^+$ and $\mu^-$ varied symmetrically with the delay time towards $\mu^+=\mu^-=1$.
This behavior demonstrates fast demagnetization of the Pt 5d magnetic moment, which has been firstly observed by the element-specific hard x-ray trXMCD technique presently developed.
The time scale of demagnetization is estimated to be $\tau_\textrm{Pt}=0.61(4)\ \textrm{ps}$ via fitting with a single exponential function as in the case of trMOKE.
The fitting result is shown by the dashed lines in Fig.~\ref{fig:fluspec}(c).
At longer delay times shown in the inset of Fig.~\ref{fig:fluspec}(c), both $\mu^+$ and $\mu^-$ recovered to the values to those before irradiation of the optical pump laser within 1 ns.
The time scale of recovery starting around 200~ps after slower recovery is estimated to be 260 ps via fitting the results using the exponential function;
this value is close to the recovery time scale of total magnetization, i.e., 200 ps, which has been determined from the trMOKE results shown in the inset of Fig.~\ref{fig:trKerr}(b).
This similarity between the time scales indicates that the Pt and total magnetic moment nearly attain an equilibrium state during the recovery process in 200-260~ps.

\begin{figure}
\begin{center}
  \includegraphics[clip,width=8cm]{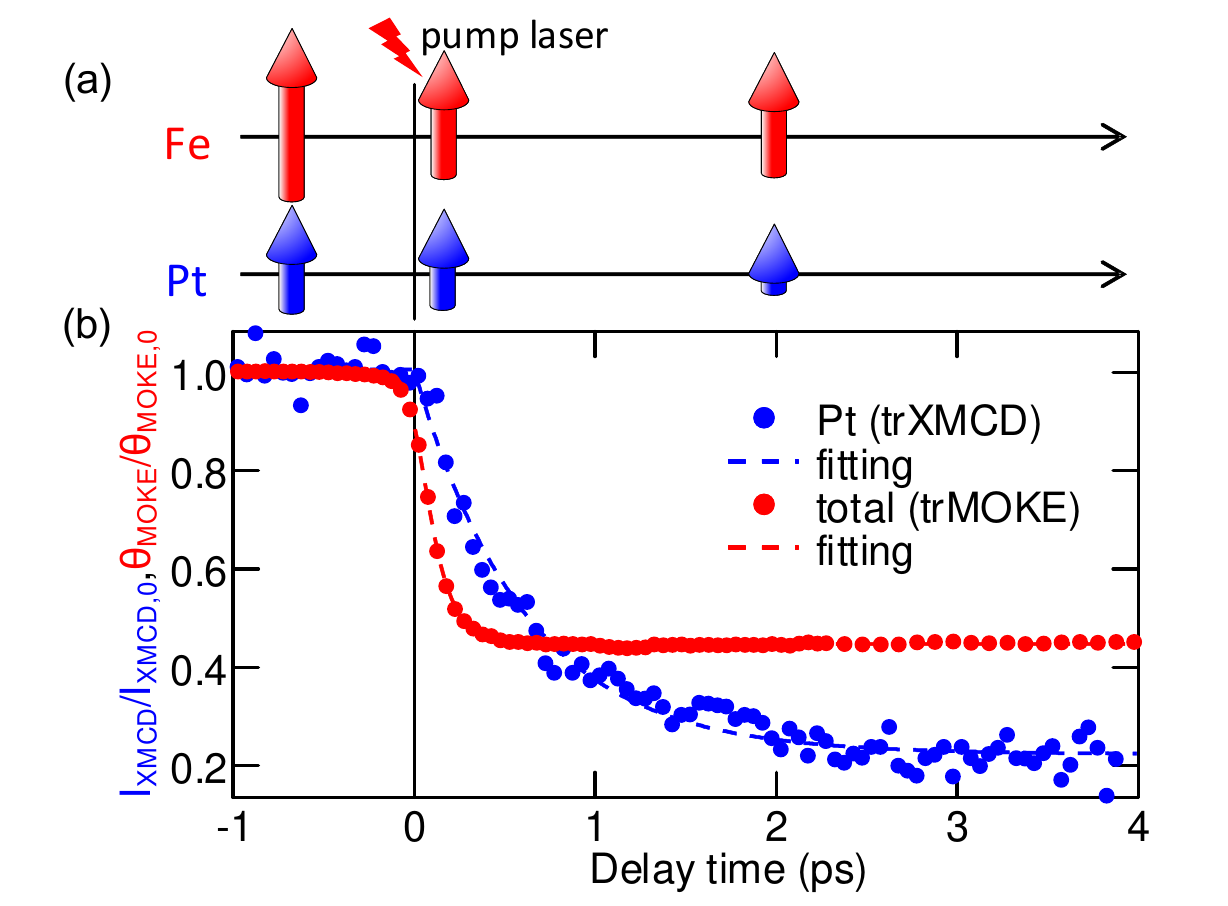}
  \caption{(a) Schematic diagram of the magnetization dynamics determined using trXMCD and trMOKE. (b) Normalized trXMCD and trMOKE intensities.
}
\label{fig:CmpXmcdKerr}
  \end{center}
\end{figure}

The magnetization dynamics of Fe and Pt are schematically depicted in Fig.~\ref{fig:CmpXmcdKerr}(a) and corresponding results of trXMCD and trMOKE are plotted together, as shown in Fig.~\ref{fig:CmpXmcdKerr}(b).
XMCD amplitudes and Kerr rotation angles have been normalized to the values unexcited state ($t<0$) to discuss the relative changes in the magnetic moment in Pt and total magnetization.
The spin and orbital magnetic moment values of Fe and Pt have been determined to be $m_\textrm{spin}^\textrm{Fe} = 2.28 \mu_\textrm{B}$, $m_\textrm{orb.}^\textrm{Fe} = 0.21 \mu_\textrm{B}$, $m_\textrm{spin}^\textrm{Pt} = 0.30 \mu_\textrm{B}$, and $m_\textrm{orb.}^\textrm{Pt} = 0.036 \mu_\textrm{B}$~\cite{Ikeda2017}.
The magnetic moment of the Fe 3d band contributes to 88\% of total magnetization.
It would be relevant to assume that trMOKE results mostly reflect the Fe magnetic moment and used to discuss ultrafast dynamics of Fe.
One can see that Fe is demagnetizing much faster than Pt, which was characterized by the different demagnetization times, $\tau_\textrm{total}=0.1\ \textrm{ps}$ and $\tau_\textrm{Pt}=0.6\ \textrm{ps}$.
In regard to the amplitude of demagnetization of Fe and Pt, Pt is more demagnetized than Fe. 
Pt magnetic moment is decreasing to 20\% of the full magnetization state, while Fe remains the magnetic moment more than 40\% of the full magnetic moment. This means that the ratio of magnetic moments of Fe and Pt changes, but the absolute value of Fe magnetic moment is still lager than Pt in the region of $0 < t < 0.6\ \textrm{ps}$, even where Fe magnetization decreases very rapidly.
Pt magnetization is induced by forming alloys with other ferromagnetic materials such as Fe and Co, and ferromagnetic coupling characterizes the magnetic nature of L1${_0}$-FePt. 
The difference of demagnetization ratio suggests photo-modulated ferromagnetic coupling or electronic hybridization between Fe and Pt.

A possible reason of the element-dependence of demagnetization time scales would be difference of the partial density of states (pDOS) in the vicinity of the Fermi level $E_\textrm{F}$.
Pt 5d bands exist in an energy region lower than Fe 3d bands~\cite{Mendil2014,Lu2010} and pDOS of Fe is larger than that of Pt around the Fermi energy.
Irradiation of the optical laser of $h\nu=1.5~\textrm{eV}$ can affect the electronic states with the energy of $E_\textrm{F}-h\nu<E<E_\textrm{F}+h\nu$ so that electron temperature of the system will increase instantaneously.
The spin temperatures of Fe could increase more rapidly than that of Pt since the pDOS in that energy region is much more than pDOS of Pt.
Furthermore, elemental Pt is a paramagnetic metal and the ferromagnetic moment of Pt in L1${}_0$-FePt is induced by the ferromagnetic element of Fe~\cite{Suzuki2005}.
By taking these results into account, the element-dependent magnetization dynamics could be explained as follows.
The laser pulse modifies the electronic temperature and causes the preferential demagnetization of Fe. 
Then the magnetic moment of Pt disappears following the demagnetization of Fe.

A previous study for the systematic trXMCD measurements of 3d and 4f pure elements and alloys indicated that demagnetization time scales are proportional to magnetic moments~\cite{Radu2015}.
However, for the present results of the 3d and 5d metal alloys, the magnetic moment of Pt is seven times smaller than Fe; nevertheless, the demagnetization time of Pt is six times longer than that of Fe.
This relatively large difference in demagnetization time scales might be related to the mechanism of AOHDS in the FePt granular.
As indicated previously, it was established that GdFeCo shows AOHDS~\cite{Lambert2014}, and in a past study of trXMCD of GdFeCo, ferrimagnetic GdFeCo also exhibited element dependent demagnetization dynamics with $\tau_\textrm{Gd}=0.43\ \textrm{ps}>\tau_\textrm{Fe}=0.1\ \textrm{ps}$, which causes the transient ferromagnetic state~\cite{Radu2011}.
Thus element-dependent time scales that were confirmed for L1${}_0$-FePt in our study and its similarity to the abovementioned GdFeCo report, indicates that element dependence of time scales is present in both ferro- or ferri- magnetic materials exhibiting AOHDS.
Further experimental and theoretical studies have to be done to understand the mechanism of demagnetization process and AOHDS in the 5d systems.

In conclusion Pt magnetization dynamics were studied in L1${}_0$-FePt thin films with trXMCD using ultrashort and circularly polarized XFEL pulses at the Pt L edge, in the hard x-ray region.
With the complementary use of trMOKE measurement, we demonstrated element-specific observation of demagnetization process with the different time scale of Pt and total magnetization as $\tau_\textrm{Pt}=0.6~\textrm{ps}$ and $\tau_\textrm{total}=0.1~\textrm{ps}$, and found the transient magnetic state, where the ratio of the magnetic moment in Pt and Fe was changed.
To obtain truly element-specific information of the Fe magnetic moment in the present system, trXMCD experiment in the similar time resolution at the Fe L edges using soft x-ray XFEL will be necessary.
In the future, our newly-developed trXMCD technique can be extended to the Pt L${}_2$ edge and transient spin and orbital magnetic moments may be obtained by sum rule analysis~\cite{Boeglin2010}.

\begin{acknowledgments}
  This work was partially supported by Toray Science and Technology Grant of Toray Research Foundation and Grant-in-Aid for Young Scientists (B) (No. 17K14334) of the Japan Society for the Promotion of Science (JSPS).
  This work was also partially supported by the Ministry of Education, Culture, Sports, Science and Technology of Japan (X-ray Free Electron Laser Priority Strategy Program).
The XFEL experiments were performed at SACLA with the approval of the Japan Synchrotron Radiation Research Institute (JASRI) (Proposal No. 2017B8060, 2017B8088).
  K.~Y. acknowledges the support from ALPS program of the University of Tokyo.
\end{acknowledgments}

\end{document}